\documentstyle[12pt,epsf]{article}

\date{}

\begin{document}

\title{\bf Soliton scattering in the $O(3)$ model on a torus}
\author{R. J. Cova \\
{\em Department of Mathematical Sciences, University of Durham}, \\
{\em Durham DH1 3LE, England} \\
{\em Departamento de F\'{\i}sica FEC, Universidad del Zulia}, \\
{\em Apartado 526, Maracaibo, Venezuela}
\and W. J. Zakrzewski \\            
{\em Department of Mathematical Sciences, University of Durham}, \\
{\em Durham DH1 3LE, England}}

\maketitle
\nonumber

\thispagestyle{empty}
\abstract
{Using numerical simulations, the stability and scattering properties of the
$O(3)$ model on a two-dimensional torus are studied. Its solitons are found
to be unstable but can be stabilized by the addition of a Skyrme-like term to
the Lagrangian.  Scattering at right angles with respect to the initial
direction of motion is observed in all cases studied. The model has no
solutions of degree one, so when a field configuration that resembles a
soliton is considered, it shrinks to become infinitely thin. A comparison of
these results with those of the model defined on the sphere is made.}

\newpage
\section{Introduction \label{sec:intro}}

The non-linear $O(3)$ model in two dimensions appears as a
low dimensional analogue of non-abelian gauge field theories in four
dimensions. This analogy relies on common properties like conformal
invariance, non-trivial topology, existence of solitons, hidden symmetries and
asymptotic freedom. Among various applications, the $O(3)$ model has
been used in the study of the quantum Hall effect and in 
models of high-$T_{c}$ superconductivity.  In solid state physics it
arises as the continuum limit of an isotropic ferromagnet, and
in differential geometry the solutions of the model are known as
harmonic maps.  

The classical $O(3)$ model defined on the sphere, or on compactified plane,
has been amply discussed in the literature \cite{lit1,lit2}. Its
finite-energy solutions are static solitons or instantons describing lumps of
energy. The model in three dimensional space-time is not integrable, and so
to study the time evolution of its solitons one must resort to numerical
simulations. The stability of the solitons has been analyzed in reference
\cite{leese1}. Due to the conformal invariance of the theory in two spatial
dimensions, these soliton-like lumps are unstable in the sense that they
change their size under any small perturbation, either explicit or introduced
by the discretization procedure.  When the lumps collide head-on, they scatter in general at 90$^{\circ}$ to the initial direction
of motion in the center-of-mass frame \cite{leese2,wjz}. Some time after the
collison the instability of the solitons makes them shrink at an
ever-increasing rate and they become so spiky that the numerical simulations
break down. However, it has been shown \cite{wjz,pey} that this instability
of the model can be cured by the addition of two extra terms to the
Lagragian. The first one resembles the term introduced by Skyrme in his
nuclear model in four dimensional space-time \cite{sky}, and the second one
is an additional potential term. The solitonic solutions of this modified
model are stable lumps (skyrmions) which repel each other and scatter at
90$^{\circ}$ when sent towards each other with sufficient speed.  

In the present paper we study the evolution properties of the $O(3)$ model
when periodic boundary conditions are imposed; this amounts to defining the
classical model on a two-dimensional torus. This approach looks more physical
than the one on the sphere in the sense that the solitons are located in a
finite volume from the outset.  In any case, a comparison between both the
toroidal and the spherical approaches is certainly of interest, if only to
check the consistency of the two results.  

In the next section we present the $O(3)$ model on the torus, and explain the
numerical set up in the following one. Solitons of degree one in both the
$O(3)$ scheme and its Skyrme version are discussed in section
\ref{sec:1soliton}, whereas their scattering is studied in section
\ref{sec:scattering}. Section \ref{sec:conclusions} completes our paper with
some conclusions.

\section{The $O(3)$ model on the torus \label{sec:torus}}

The non-linear $O(3)$ model involves three real scalar fields
\( \vec{\phi}(x^{\mu})\equiv \{\phi_{a}(x^{\mu}), a=1,2,3\}\, \)
with the constraint that for all 
\( x^{\mu} \equiv (x^{0},x^{1},x^{2})=(t,x,y) \) 
the fields lie on the unit sphere $S_{2}$:
%
\begin{equation}
\vec{\phi}.\vec{\phi}=1.
  \label{eq:constraint}
\end{equation}
Subject to the above constraint the Lagrangian density and the 
corresponding equations of motion read
%
\begin{equation}
{\cal L}=\frac{1}{4}
(\partial_{\mu}\vec{\phi}).(\partial^{\mu}\vec{\phi}),
  \label{eq:lagrangian}
\end{equation}
\begin{equation}
\partial^{\mu}\partial_{\mu}\vec{\phi}+
(\partial^{\mu}\vec{\phi}.\partial_{\mu}\vec{\phi})\vec{\phi}=\vec{0}.
\label{eq:motion}
\end{equation}
For any value of $t$, 
the fields $\vec{\phi}$ are mappings from the torus $T_{2}$ to
the sphere $S_{2}$, {\em i.e.,} they satisfy the periodic boundary conditions
%
\begin{equation}
\vec{\phi}(x+mL,y+nL)=\vec{\phi}(x,y),
  \label{eq:boundaryphi}
\end{equation}
where $m,n=0,1,2,...,$ and the period $L$ denotes the size of 
the square torus. 

It is convenient to describe the model in terms of one independent 
complex field $W$, related to $\vec{\phi}$ via
%
\begin{equation}
\vec{\phi}=(\frac{W+\bar{W}}{|W|^{2}+1},
i\frac{-W+\bar{W}}{|W|^{2}+1},\frac{|W|^{2}-1}{|W|^{2}+1}). 
  \label{eq:projection}  
\end{equation}
Introducing complex coordinates $z=x+iy$ and $\bar{z}=x-iy$ on
the torus and using the handy notation $\partial_{z}W=W_{z}, \,
\partial_{z}(\partial_{\bar{z}}W)=W_{z\bar{z}}$, {\em etc.}, 
the equation of motion (\ref{eq:motion}) for the static field 
configurations becomes
%
\begin{equation}
W_{z\bar{z}}-\frac{2\bar{W}W_{z}W_{\bar{z}}}{|W|^{2}+1}=0,
\label{eq:motionw}
\end{equation}
whereas the boundary conditions (\ref{eq:boundaryphi}) take the form
\begin{equation}
W(z+mL+inL)=W(z).
  \label{eq:boundaryw}
\end{equation}
Whence, our static solitons or instantons are elliptic functions
that may be expressed as \cite{mcgraw}
%
\begin{equation}
W=\lambda \prod_{j=1}^{\kappa}
\frac{\sigma(z-a_{j})}{\sigma(z-b_{j})}, 
\label{eq:instanton}
\end{equation}
with the zeros $(a_{j})$ and poles $(b_{j})$ subject to the selection
rule
%
\begin{equation}
\sum_{j=1}^{\kappa} a_{j}=\sum_{j=1}^{\kappa} b_{j}.
  \label{eq:rule}
\end{equation}
The complex number $\lambda$ is related to the size of the soliton and
$\kappa$ is the order of the elliptic function $W$.
The Weierstrass $\sigma$-function is defined on $T_{2}$ and satisfies
the pseudo-periodicity property \cite{mcgraw}
%
\begin{equation}
\sigma (z+mL+inL)=(-1)^{(m+n+mn)}
\exp (\frac{\pi}{L}(m-in)[z+\frac{1}{2}(m+in)L])
\sigma (z).
\label{eq:periodicity}
\end{equation}
For a square torus we have the so-called lemniscatic case,
where $\sigma$ posseses a Laurent expansion of the form
%
\begin{equation}
\sigma (z)=\sum_{j=1}^{\infty}{c}_{j}z^{4j+1}, 
\label{eq:expansion}
\end{equation}
where the real coefficients ${c}_{j}$ depend on $L$.

Note that the periodicity of the torus means that our study can be
simplified to the consideration of the system in a 
\underline{fundamental cell} delimited by the vertices
\begin{equation}
(0,0),\, (L,0),\, (L,L),\, (0,L).
\label{eq:cell}
\end{equation}

We are interested in static finite-energy solutions, which in the language of
differential geometry are harmonic maps \( T_{2} \rightarrow S_{2} \). These
maps have been extensively studied in differential geometry
\cite{eells1,eells2}. They are partitioned into homotopy sectors parametrized
by an invariant integral index $Q$, the degree of the map, defined as usual by
taking a two-form from $S_{2}$ to $T_{2}$ via the pull back map
$W^{*}$. For a given map \( W:T_{2} \rightarrow S_{2} \) we can pull back the
Kahler form
%
\begin{equation}
\omega=4i \frac{d\xi \wedge d\bar{\xi}}{(1+\xi\bar\xi)^{2}}, 
\qquad \xi \, \epsilon \, S_{2}
\label{eq:form},
\end{equation}
and define 
%
\begin{equation}
Q=\frac{1}{C} \int_{T_{2}} W^{*}\omega,
\label{eq:Qdefinition}
\end{equation}
where the constant $C$ normalizes $Q$ to an integer. 
Expanding (\ref{eq:form}) in terms of $z$ and setting $C=-8\pi$ 
we obtain
%
\begin{equation}
Q=\frac{1}{\pi}\int_{T_{2}} \frac{|W_{z}|^{2}-|W_{\bar{z}}|^{2}}
{(|W|^{2}+1)^{2}}\, dxdy.
\label{eq:chargew}
\end{equation}
The potential energy $V$, as derived from the Lagrangian
(\ref{eq:lagrangian}), and the topological index satisfy
%
\begin{equation}
V \geq 2\pi |Q|.
\label{eq:bound}
\end{equation}
The instanton-solutions correspond to the equality in (\ref{eq:bound}):
Solutions carrying $Q>0$ $(Q<0)$ imply
$W_{\bar{z}}=0$ $(W_{z}=0)$, which are the Cauchy-Riemann
conditions for $W$ being an analytic function of $z$ $(\bar{z})$.

The periodic solitons (\ref{eq:instanton}) have been studied in a variety of
contexts. In reference \cite{richard}, for example, they were used to compute
the contribution of instantons to the partition function.   

\section{Numerical procedure for the time evolution \label{sec:numerical}}

So far we have discussed the static field configurations. Now we concentrate
on their dynamics, paying particular attention to their stability,
scattering properties, {\em etc..} As our model is not integrable, the study
of the evolution of our fields requires numerical techniques. Hence, we treat
configurations (\ref{eq:instanton})-(\ref{eq:rule}) as initial conditions for
our evolution, studied numerically.  The field $\vec{\phi}$ is evolved
according to the equation of motion (\ref{eq:motion}). We compute the series
(\ref{eq:expansion}) up to the fifth term, the coefficients ${c}_{j}$ being
in our case negligibly small for $j \geq 6$. The numerical set up is similar
to that used in our previous papers (see \cite{me} for example): We employ
the fourth-order Runge-Kutta method and approximate the spatial derivatives
by finite differences. The Laplacian is evaluated using the standard
nine-point formula and, to further check our results, a 13-point recipe is
also utilized.  We work on a 200x200 periodic lattice ($n_{x}=n_{y}=200$)
with spatial and time steps $\delta x$=$\delta y$=0.02 and $\delta t$=0.005,
respectively. The size of our torus is then $L=n_{x}\times\delta x=4$.  

Unavoidable numerical truncation errors introduced at various \linebreak
stages of the calculations gradually shift the 
fields away from the unit sphere (\ref{eq:constraint}). So we rescale
$$
\vec{\phi} \rightarrow \vec{\phi}/\sqrt{\vec{\phi}.\vec{\phi}}
$$
every few iterations. The error associated with this procedure 
is of the order of the accuracy of our calculations. Each time,
just before the rescaling operation, we evaluate the quantity
\( \mu \equiv \vec{\phi}.\vec{\phi} - 1 \)\,
at each lattice point. Treating the maximum of the
absolute value of $\mu$ as a measure of the
numerical errors, we find that max$|\mu|$ $\approx$ 10$^{-8}$.
This magnitude is useful as a guide to determine how
reliable a given numerical result is. Usage of an unsound 
numerical procedure like, say, taking 
$\delta x$ $<$ $\delta t$ in the Runge-Kutta evolution, shows 
itself as a rapid growth of max$|\mu|$; such increase
also occurs when the solitons become infinitely spiky.

We use the global $U(1)$ symmetry of (\ref{eq:instanton}) to choose
$\lambda$ real; the value $\lambda$=1 has been used in all our
simulations. 

\section{Solitons of degree one \label{sec:1soliton}}
\subsection{O(3) case}

From the theory of elliptic functions we know that the simplest non-trivial
elliptic functions are of order two. This implies that the $O(3)$ model on the
torus possesses no single-soliton solutions. This fact may also be understood
in the context of differential geometry \footnote{We thank J.M. Speight for
showing us a simple mathematical proof of this fact.}: The harmonic maps  \( M
\rightarrow S_{2} \) ($M$ an orientable surface) have holomorphic
representatives (instantons) of any degree provided that it is greater than
the genus of $M$ \cite{eells1,eells2}. Clearly, for $M=T_{2}$ the index of the
maps must be greater than unity. Note that the degree $Q$ in
(\ref{eq:chargew}) is numerically equal to the order $\kappa$ of
(\ref{eq:instanton}) only when the latter is greater than one. Thus,
an order-one solution (the trivial solution) carries degree zero, not one.  

In order to study a single soliton on $T_{2}$, we ignore the selection 
rule (\ref{eq:rule}) and take the configuration
%
\begin{equation}
W_{1}=\frac{\sigma(z-a)}{\sigma(z-b)}, 
\qquad a \ne b,
\label{eq:w1}
\end{equation}
which describes a quasi-periodic lump that instead of (\ref{eq:boundaryw}) 
satisfies 
%
\begin{equation}
W_{1}(z+mL+inL)=
\exp [\frac{\pi}{L}(m-in)(b-a)]W_{1}(z).
\label{eq:periodicityw1}
\end{equation}
But a periodic solution may be constructed by taking a field whose
values in the \underline{sub-cell} of vertices
$$
(l,l),\, (L-l,l),\, (L-l,L-l),\, (l,L-l), \qquad l \ll L,
$$
are given by $W_{1}$ and in the rest of the fundamental cell (\ref{eq:cell})
are given by a suitably chosen interpolating function. 

So let us periodize $W_{1}$ along the $x$-axis with the help 
of the {\em ansatz}
\begin{equation}
W_{h}=A(y)\tanh[\alpha(x-L)] + B(y); \quad x\, \epsilon\, [L-l,L+l], 
                                         \,  y\, \epsilon\, [0,L].
\label{eq:wh}
\end{equation}
(note that \( [L-l,L+l]=[0,l] \cup [L-l,L] \)). The complex functions 
$A(y)$ and $B(y)$ are obtained by demanding periodicity and continuity 
of (\ref{eq:w1}) and (\ref{eq:wh}). One deduces 
\begin{equation}
\left.
\begin{array}{lll}
A(y)&=&[W_{1}(l+iy)-W_{1}(L-l+iy)]/2\tanh(\alpha l),\\
B(y)&=&[W_{1}(l+iy)+W_{1}(L-l+iy)]/2.
\end{array}
\right.
\label{eq:AB}
\end{equation}
Therefore, our horizontally-periodic configuration is
%
\begin{equation}
W_{H}(x,y)=
\left\{
\begin{array}{lll}
 W_{1}; & x \, \epsilon \, [l,L-l],   & y \, \epsilon \, [0,L];
  \\  
 W_{h}; & x \, \epsilon \, [L-l,L+l], & \; y \, \epsilon \, [0,L].
\end{array}
\right.
\label{eq:wH}
\end{equation}

A similar periodization is now performed on $W_{H}$ along the vertical
axis. It turns out that the field $W_{p}$ thus obtained is periodic in 
both $x$ and $y$. It has the appearance
\newpage
%
\begin{equation}
W_{p} =
\left\{
\begin{array}{lll}
  W_{H}; & x \, \epsilon\ \, [0,L], & y \, \epsilon\, [l,L-l]; 
  \\
  C(x)\tanh[\alpha(y-L)]+D(x);
         & x \,\epsilon\ [0,L],     & y \, \epsilon \, [L-l,L+l],
\end{array}
\right.
\label{eq:wp}
\end{equation}
with
\begin{equation}
\left.
\begin{array}{lll}
C(x)&=&\{W_{H}(x+il)-W_{H}[x+i(L-l)]\}/2\tanh(\alpha l),\\
D(x)&=&\{W_{H}(x+il)+W_{H}[x+i(L-l)]\}/2.
\end{array}
\right.
\label{eq:CD}
\end{equation}
For the length $l$ we may take ten lattice points, so that $l=0.2 \ll L=4$.
We set the value of $\alpha$  equal to 20, and for the zero and pole of 
(\ref{eq:w1}) we elect  
\begin{equation}
a=(2.05,1.75), \quad b=(1.95,2.25).
\label{ab1lump}
\end{equation}
We have numerically checked that the {\em ansatz} (\ref{eq:wp}) has $Q=1$,
and so it may be regarded as a map \( T_{2} \rightarrow S_{2} \) 
of degree one. 

\begin{figure}
\epsfverbosetrue
\epsfysize=10cm
\centerline{\epsfbox{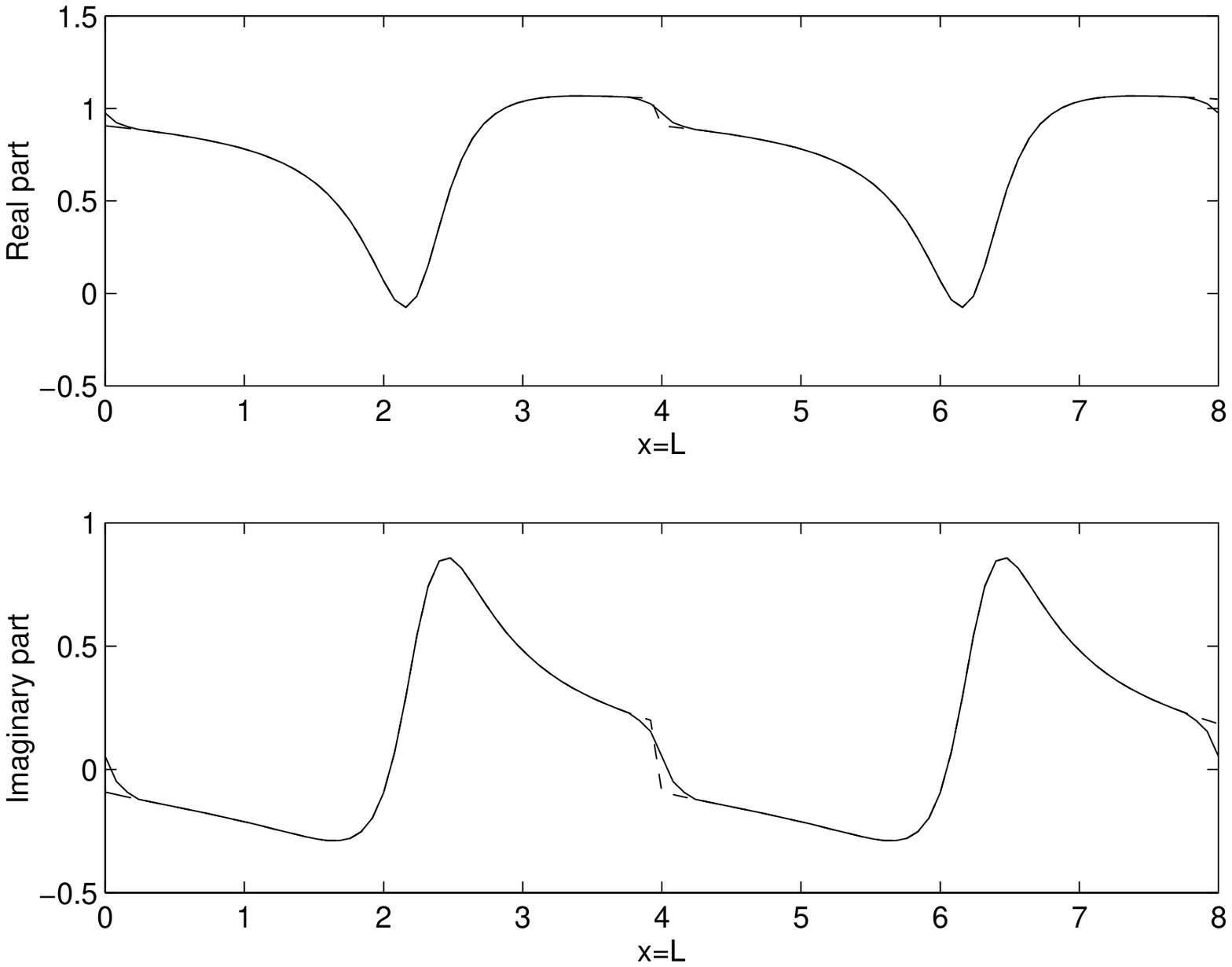}}
\epsfysize=10cm
\centerline{\epsfbox{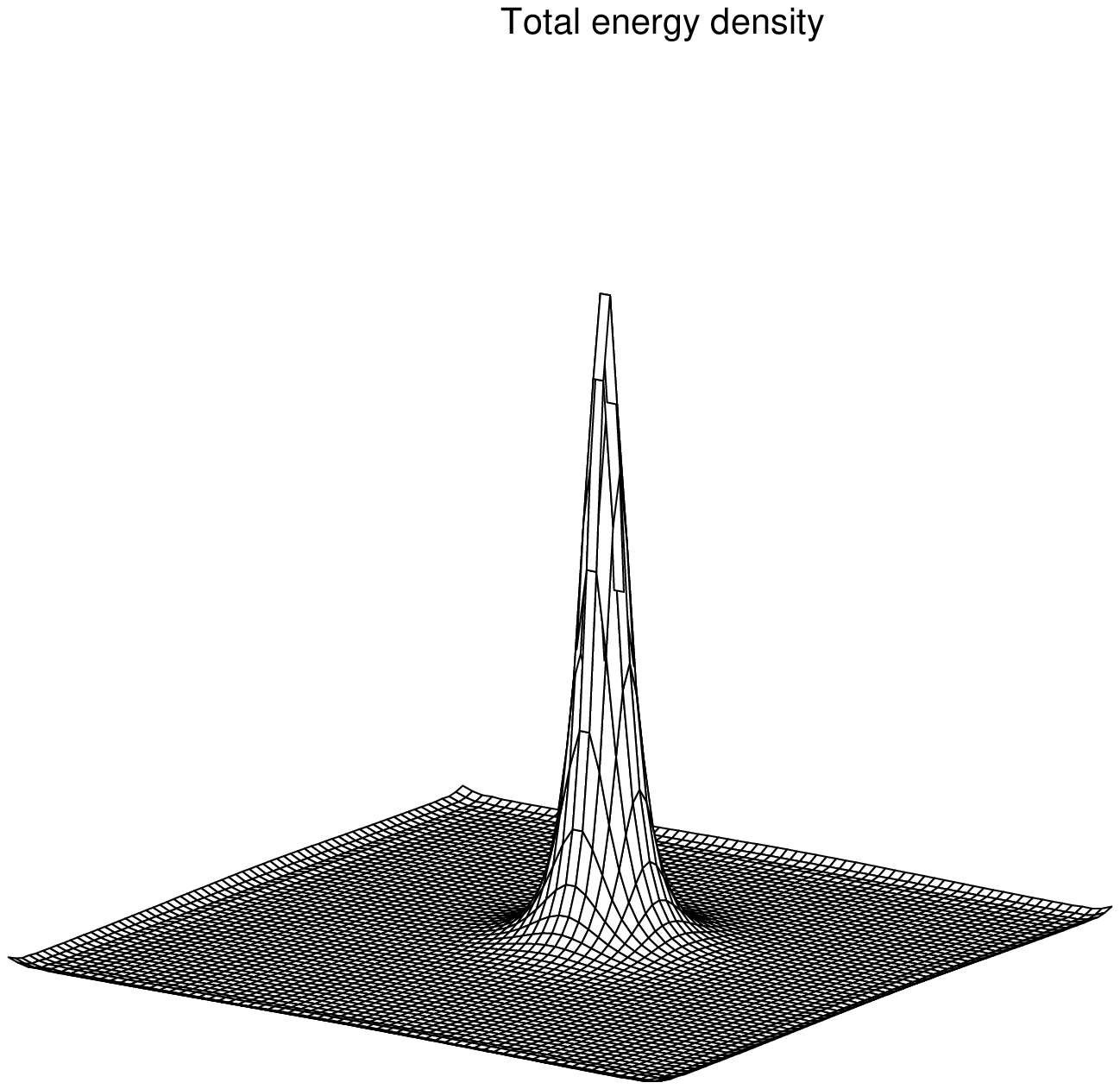}}
\caption{\underline{Above}: $W_{1}$ (dashed line) and its periodized
version $W_{p}$ (solid line) along the line $y=2$ of the fundamental
cell. \underline{Below}: A typical picture for the total energy density 
associated with $W_{p}$ at $t=0$. The folds at the edges are caused 
by the periodization procedure.}
\label{fig:wpline}
\end{figure}

The upper half of figure (\ref{fig:wpline}) illustrates the 
periodization of $W_{1}$ for a representative line of 
the fundamental cell. The lower half exhibits the
total energy density associated with the periodized field 
$W_{p}$.
It is apparent from this picture that the periodization procedure
introduces some perturbation at the borders of the grid in 
the form of small folds. Under the  numerical evolution these
perturbations propagate towards the center of the lattice
and collapse the lump. In order to minimize such an effect, we improve the
initial conditions by ironing out the folds. We
do this by implementing a damping function, $\gamma$, that 
rescales 
\begin{math} \partial_{t}\vec{\phi_{p}} \rightarrow 
\gamma \partial_{t}\vec{\phi_{p}\/}, \, \gamma \le 1.
\end{math}
\,
The absorbtion is switched off at the time ($t_{0}$)
when the folds have disappeared; the resultant configuration 
serves us as a better, improved set of initial conditions. 

\begin{figure}
\epsfverbosetrue
\centerline{\epsfbox{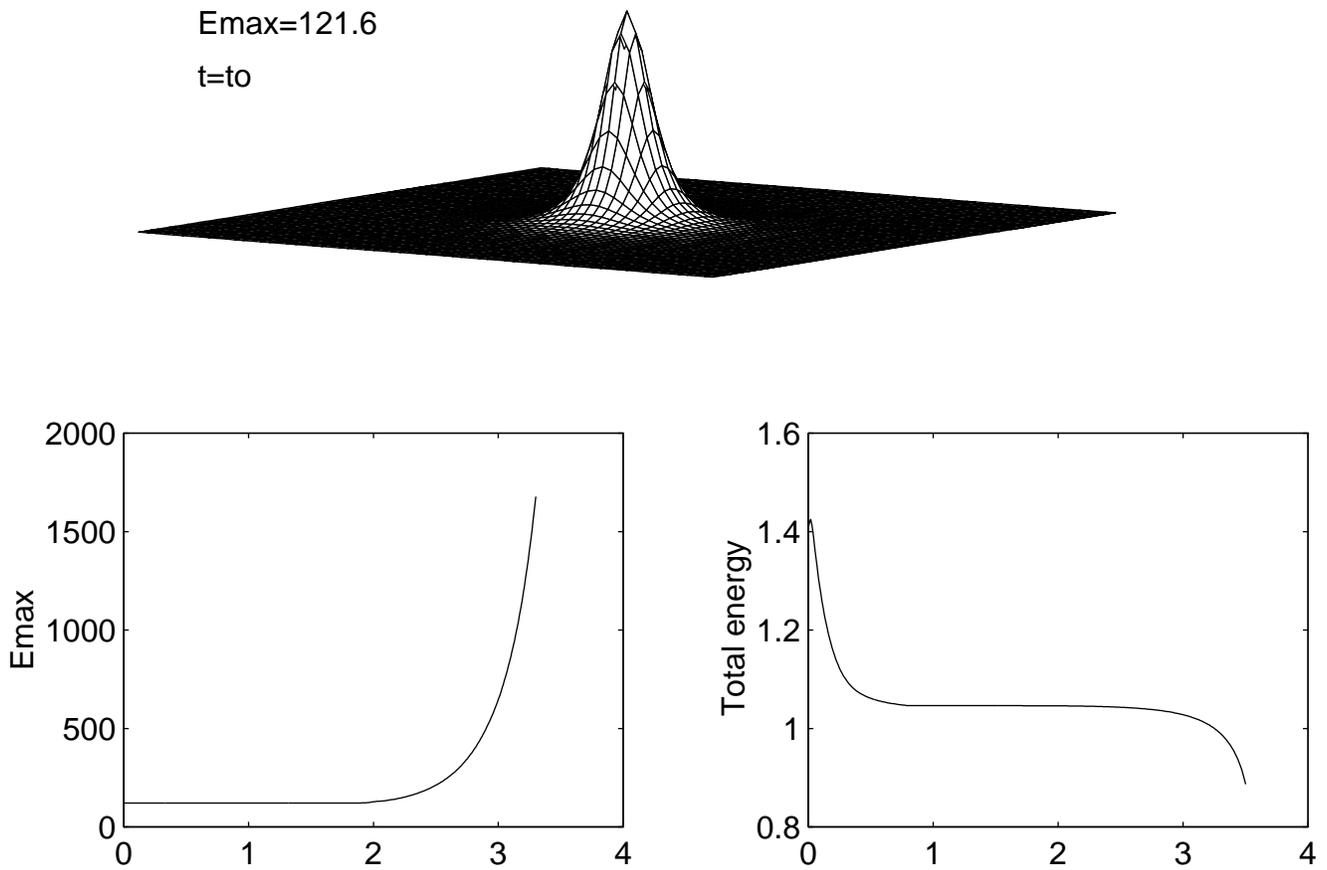}}
\caption{\underline{Above}: Total energy density at $t_{0}$=0.8, 
corresponding to our prepared, improved initial one soliton-like 
configuration. \underline{Below}: The maximum value of the total 
energy density ($Emax$) and the total energy {\em vs.} $t$.
The lump grows  infinitely tall soon after $t_{f}\approx$ 3.5, and the 
numerical procedure collapses.}  
\label{fig:eflat}
\end{figure}

During the preparatory stage the total energy undergoes a small decrease, in
conformity with the absorption that is taking place.  Once the latter is
turned off, the energy settles near the expected value of one and remains
constant until the time ($t_{f}$) when the total energy density becomes so
spiky that the numerical procedure breaks down [see figure \ref{fig:eflat}]. 
Moreover, we have checked that these results do not depend on how the initial
conditions were prepared, nor on whether a 9-point or a 13-point laplacian
operator was used in the simulations.  Having performed many such simulations
we are convinced that our results are genuine, {\em i.e.,} the shrinking is
genuine and not a numerical artifact.

\subsection{Skyrme case}

Next we look at possible ways to stabilizing our solitons. Guided by the
experience with the $O(3)$ model in the plane (where the shrinking can be
prevented by the addition of two extra terms to the Lagrangian -the
Skyrme and the potential term-) we consider the possibility of adding the
Skyrme term alone.  Adding such a term  \cite{azca} to the $O(3)$ Lagrangian
we have: 
%
\begin{eqnarray}
{\cal L}_{skyrme}&=&\frac{1}{4}
(\partial_{\mu}\vec{\phi}).(\partial^{\mu}\vec{\phi}) \nonumber \\
&-&\frac{\theta_{1}}{4}
[(\partial^{\mu}\vec{\phi}.\partial_{\mu}\vec{\phi})^{2}-
(\partial^{\mu}\vec{\phi}.\partial^{\nu}\vec{\phi})
(\partial_{\mu}\vec{\phi}.\partial_{\nu}\vec{\phi})]
\label{eq:lagsky}
\end{eqnarray}
or, in the simpler $CP^{1}$ formulation 
\begin{equation}
{\cal L}_{skyrme}= 
\frac{|W_{t}|^{2}-2|W_{z}|^{2}}{(1+|W|^{2})^{2}}
  +8\theta_{1}\frac{|W_{z}|^{2}}{(1+|W|^{2})^{4}}
     (|W_{t}|^{2}-|W_{z}|^{2}).   
\label{eq:lagskyw}
\end{equation}
The associated {\em static} equation of motion reads
%
\begin{eqnarray}
\lefteqn{0=W_{z\bar{z}}-\frac{2\bar{W}W_{z}W_{\bar{z}}}{|W|^{2}+1}} 
\nonumber \\
 & &
+\frac{4\theta_{1}}{(|W|^{2}+1)^{2}}
[2\bar{W}_{z\bar{z}}W_{z}W_{\bar{z}}-
\bar{W}_{zz}(W_{z})^{2}- 
\bar{W}_{\bar{z}\bar{z}}(W_{\bar{z}})^{2}+
W_{zz}\bar{W}_{z}W_{\bar{z}} \nonumber \\
 & & 
+W_{\bar{z}\bar{z}}\bar{W}_{\bar{z}}W_{z}
-W_{z\bar{z}}(|W_{z}|^{2}+|W_{\bar{z}}|^{2})
\nonumber \\
 & &
+\frac{2W}{|W|^{2}+1}(|W_{z}|^{2}-|W_{\bar{z}}|^{2})^{2}].
\label{eq:skyrme}
\end{eqnarray}

\begin{figure}
\epsfverbosetrue
\centerline{\epsfbox{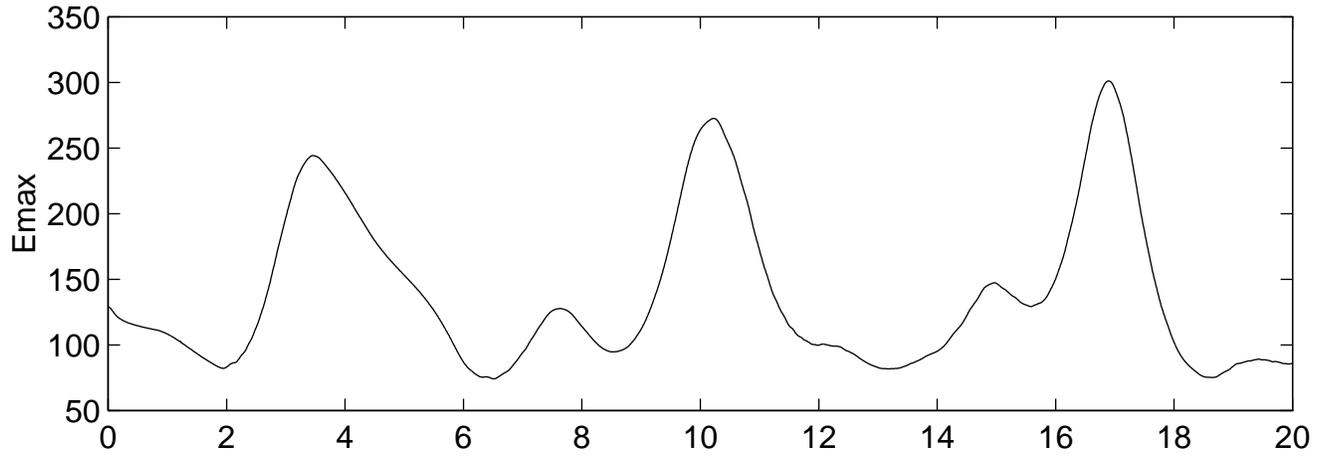}}
\caption{Peak of the total energy 
density {\em vs.} time corresponding to a single-lump in the modified 
$O(3)$ model with $\theta_{1}=0.001$. The soliton is now stable.}
\label{fig:skymod} 
\end{figure}

Indeed, we find that thanks to the extra term the energy density
of the lump does
not increase indefinitely, but instead it oscillates with time in a stable
manner. In figure (\ref{fig:skymod}) we show the evolution of the amplitude
of the total energy density for $\theta_{1}=0.001$. Qualitatively similar
pictures are obtained for values of $\theta_{1}$ as small as $\approx$
0.00015; for smaller values $W_{p}$ is no longer stable. 

Note that $W_{p}$ does not exactly satisfy the equation of motion
(\ref{eq:skyrme}), for the term
$$
\frac{4\theta_{1}}{(|W|^{2}+1)^{2}}[\bar{W}_{zz}(W_{z})^{2}
-2\frac{W|W_{z}|^{4}}{|W|^{2}+1}]
$$
does not vanish. Nevertheless, the smallness of 
$\theta_{1}$ means that our Skyrme model is only a slight perturbation
of $O(3)$, and hence $W_{p}$ is a good, if approximate, 
solution. 

Worthy of remark is the fact that (\ref{eq:lagsky}) does not require a
potential-like term to stabilize the lumps. In the $O(3)$ system in the plane
with fixed boundary conditions, by contrast, such a term was needed to prevent
the solitons from expanding.

\section{Solitons of degree 2 \label{sec:scattering}}
%
\subsection{$O(3)$ case}
We now move on to the interesting question of collisions, limiting 
ourselves to two solitons. It is important to bear in mind that the 
preparatory stage devised for the pathological single-soliton case of section 
\ref{sec:1soliton} is not required for lumps of degree $\ge 2$.
The initial field is given by an order-two fuction of the form 
(\ref{eq:instanton})-(\ref{eq:rule}):
%
\begin{equation}
W_{2}=\frac{\sigma(z-a_{1})}{\sigma(z-b_{1})}
\frac{\sigma(z-a_{2})}{\sigma(z-b_{2})}, \quad a_{1}+a_{2}=b_{1}+b_{2}.   
\label{eq:2soliton}
\end{equation}

First, consider the situation when the solitons are symmetrically
positioned along the horizontal axis and boosted towards each other
with relative velocity $v=(0.2,0)$. We select the zeros and poles to be: 
\begin{equation}
\left.
\begin{array}{lll}
a_{1}&=&(0.77,1.95), \quad a_{2}=(3.25, 1.95); \\
b_{1}&=&(1.32, 1.95), \quad b_{2}=(2.70, 1.95).
\end{array}
\right.
\label{eq:abhorizontal}
\end{equation}
The solitons gradually shrink and then undergo a gradual expansion as they
approach each other. They collide at the centre of the grid and merge
into a complicated ringish structure, where they are no longer
distinguishable. After this process the solitons get narrower and narrower
as they re-emerge at right angles to the initial direction of motion. Due
to their instability, the shrinking process goes on until
the solitons get so spiky that the numerical procedure is no longer reliable;
this occurs for $t \approx 7$, when max$|\mu|$ as defined in section
\ref{sec:numerical} reaches $\approx$ 10$^{-4}$ and higher.  

\begin{figure}
\epsfverbosetrue
\centerline{\epsfbox{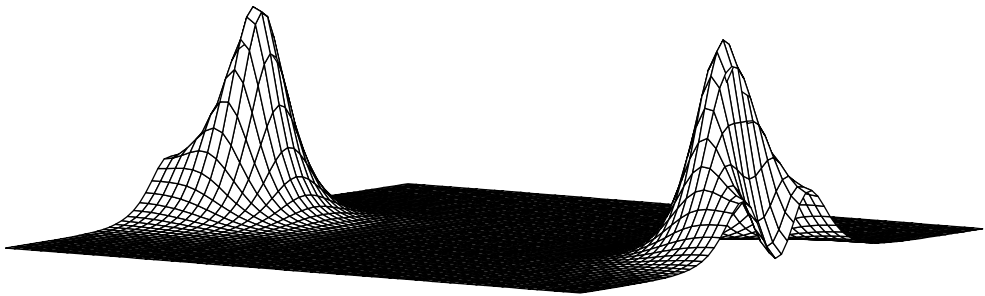}}
\caption{Total energy density corresponding
to $O(3)$ solitons moving away from the centre with
$v=(0.2,0)$. The amplitude of the lumps gradually decreases as they
approach each other, reaching a maximum when they coalesce.}
\label{fig:away}
\end{figure}

A numerically interesting feature of the periodic $O(3)$ model is that the
scattering can also be observed when the solitons are sped `away' from each
other, towards the borders of the fundamental cell. This is a good way to test
the correctness of our periodic lattice. Applying this to the solitons defined
by (\ref{eq:abhorizontal}) we also observe the scattering at 90$^{\circ}$. A
representation of this process can be viewed in figure (\ref{fig:away}).

\begin{figure}
\epsfverbosetrue
\centerline{\epsfbox{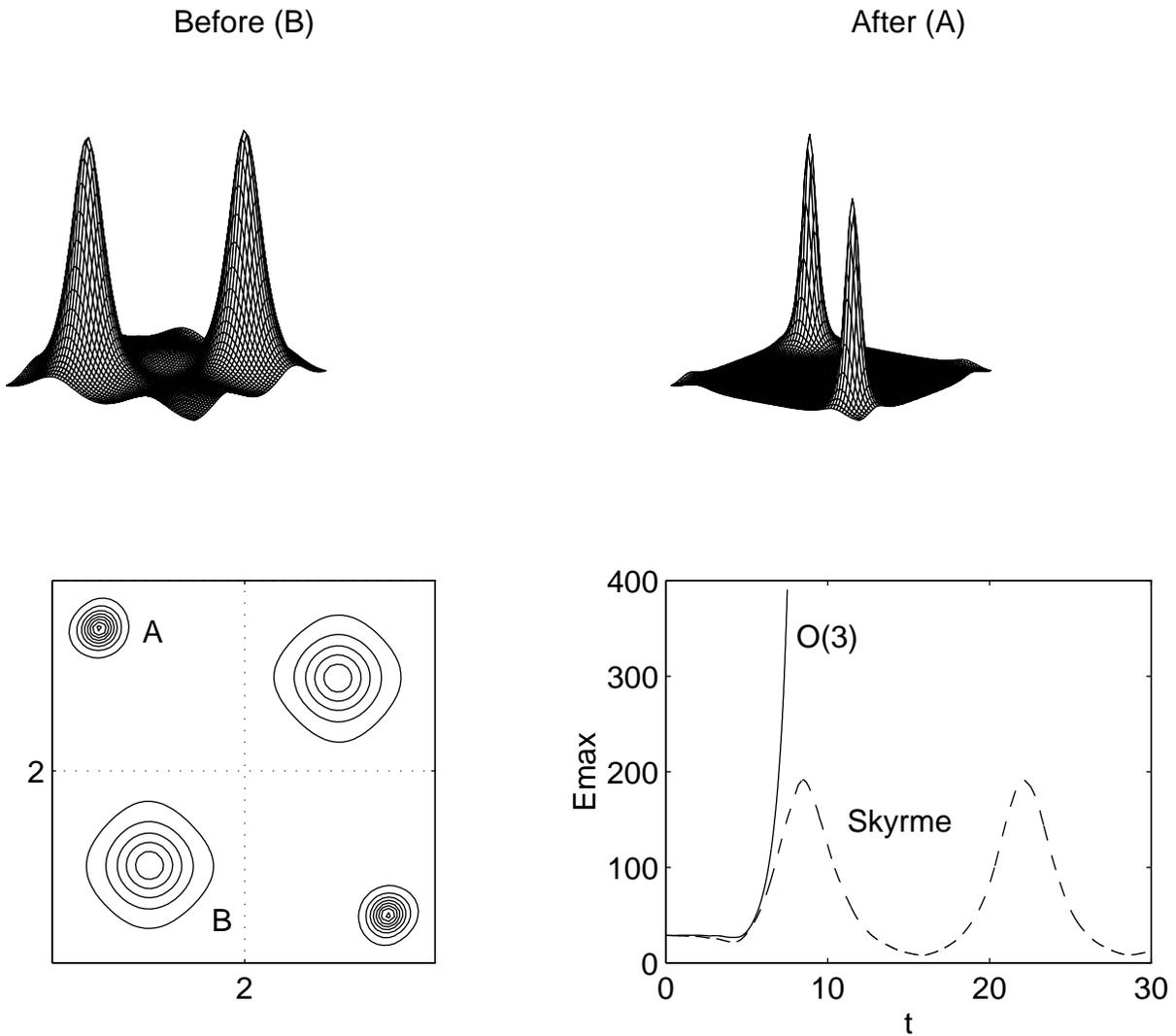}}
\caption{$O(3)$ solitons moving away from the centre along the (0,0)-(4,4)
diagonal (B). They collide at the corners and scatter at 
right angles (A). When the model is supplemented by a Skyrme term
the lumps are stable, as shown in the accompanying graph $Emax(t)$.}
\label{fig:awayo3diag}
\end{figure}

A typical head-on collision with the solitons initially placed along
a diagonal is illustrated in figure (\ref{fig:awayo3diag}). The initial 
position is achieved by the arrangement
\begin{equation}
\left.
\begin{array}{lll}
a_{1}&=&(0.95,0.75), \quad a_{2}=(3.05, 3.25); \\
b_{1}&=&(1.22, 1.95), \quad b_{2}=(2.78, 2.05).
\end{array}
\right.
\label{eq:abdiagonal}
\end{equation}

After boosting the solitons away from the centre with initial velocity
$v=\frac{\sqrt{2}}{10}(1,1)$ \,-$|v|$=0.2-, the lumps collide at the corner
(0,0)=(4,4) and re-appear from (0,4)=(4,0) at right angles to the initial
direction of motion. Of course, all four corners are nothing but the same
point; there the lumps meet, coalesce and scatter off as already explained. 
Shortly afterwards, the instability of the system manifests itself in the usual
manner, as reflected by the $O(3)$ curve in the graph $Emax(t)$ in figure
(\ref{fig:awayo3diag}). This diagram also includes the resulting curve of the
Skyrme version, as described in subsection (\ref{sub:sky2}) below. Also, when
situated in an arbitrary, non-symmetrical way within the fundamental lattice,
the solitons always scatter at ninety degrees when sent head-on against one
another (we discuss in more detail this situation in the next section).  

We may interpret the instability of (\ref{eq:2soliton}) under numerical
simulations as follows: The solitons start off satisfying the selection rule
\( a_{1}+a_{2}=b_{1}+b_{2} \), which links them in some manner. Due to
inevitable round-off errors during the numerical simulation, the
field gets perturbed and so is only approximately described by the original
field configuration.  As the perturbation is quite small it will excite mainly
the degrees of freedom which are zero modes of the original configuration.
Thus, in particular, $a_{j}$ and $b_{j}$ will start evolving but in order to
remain close to the original configuration they will keep the constraint
unbroken. Such evolution may lead to $a_{j}$ and $b{j}$, pairwise, coming
close together. This corresponds to the solitons shrinking. To see this note
that \( |a_{j}-b_{j}|/2 \) determines the size of the $j$-th soliton.  Note
that this shrinking is essentially of the same type as the well known
shrinking of any number of solitons on the sphere. We would like to stress
that since analytical solutions exist in all topological sectors of index
$\ge$ 2, this lack of stability of our two-soliton system is of a different
nature than the instability of the single-soliton configuration (and so
non-existence of a one-soliton static solution)  discussed in the previous
section.  There the solution does not exist on the lattice or in the
continuum; here the solutions do exist in the continuum but are unstable and
putting them on the lattice introduces a perturbation which sets off the
instability.  
  
\subsection{Skyrme case \label{sub:sky2}}

\begin{figure}
\epsfverbosetrue
\centerline{\epsfbox{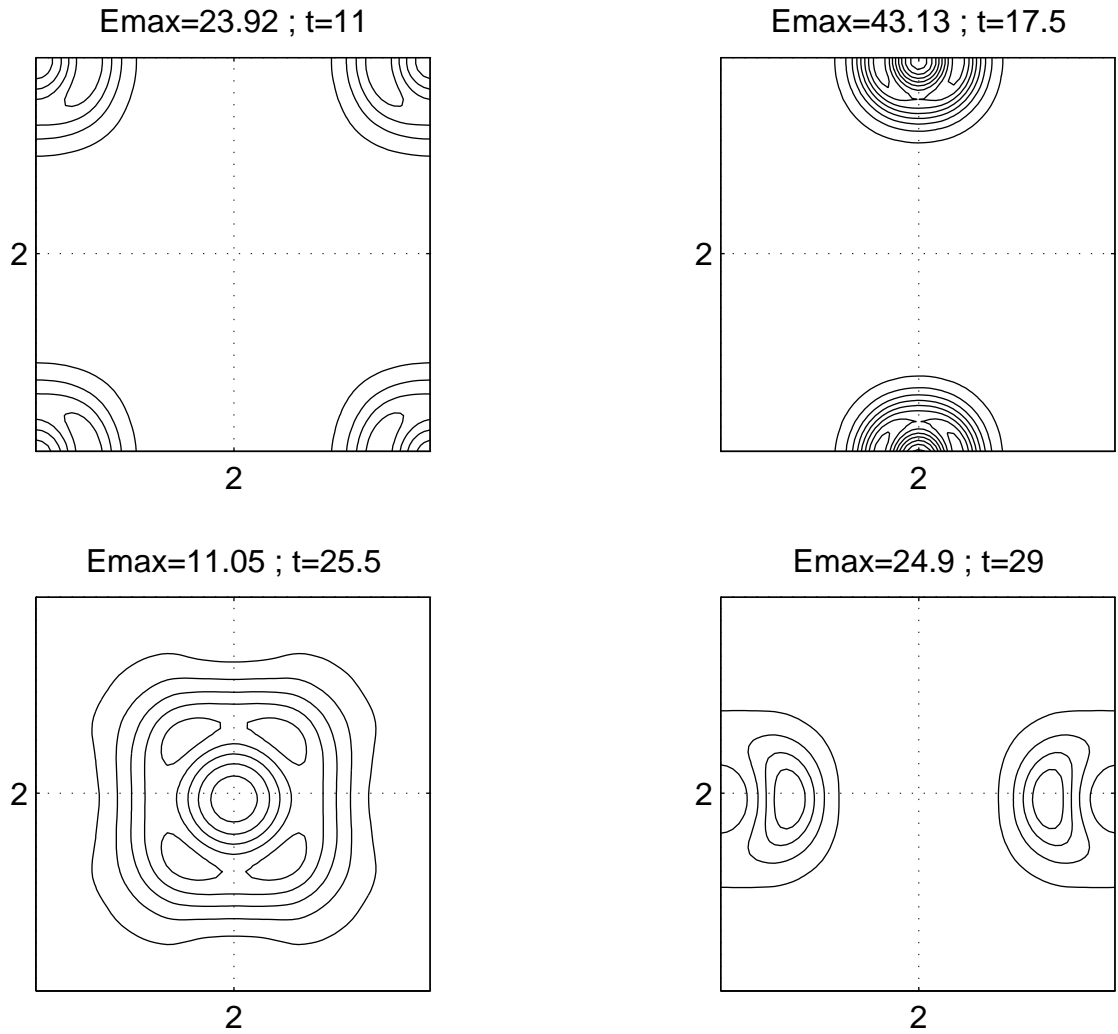}}  
\caption{Skyrmion scattering ($\theta_{1}$=1/2000) for $v=(0.2,0)$.
After scattering like the $O(3)$ solitons of figure 4, the skyrmions do 
not collapse but go on to collide at
$t$=11,\,17.5,\,25.5 and so forth. At every occasion they scatter at right
angles. This cycle repeats itself indefinitely.}
\label{fig:conaway} 
\end{figure}

Let us now consider the Skyrme Lagrangian (\ref{eq:lagskyw}) as applied to two
solitons.  Head-on collisions along the horizontal axis corresponding to the
set up (\ref{eq:abhorizontal}) proceed as in the pure $O(3)$ scheme. The
Skyrme term, however, prevents the lumps from shrinking indefinitely and
renders them stable; their motion can now be followed for as long as desired. 
For instance, the skyrmions proceed as in figure (\ref{fig:away}) but, after
90$^{\circ}$ scattering at the lattice point (0,2)=(4,2), they continue their
journey and collide thrice more, reach again their $t$=0 positions and proceed
to repeat this cycle anew, as suggested by figure (\ref{fig:conaway}). Note
the coalescence of the lumps in the corners (subplot $t=11$) which, as
mentioned before, are nothing but one and the same point. The plot of the
corresponding $Emax(t)$  -not shown- is very much like the one drawn in figure
(\ref{fig:awayo3diag}) [dashed curve].  All two-skyrmion cases shown in this
paper correspond to $\theta_{1}=1/2000$, but the same qualitative behaviour is
found for values down to $\approx$ 0.00007. Smaller values cannot prevent the
lumps from getting too thin, leading to the breakdown of our code.

An example of solitons located at two arbitrary cell points is
given by the parameters
\begin{equation}
\left.
\begin{array}{lll}
a_{1}&=&(0.77,1.30), \quad a_{2}=(3.25, 2.70); \\
b_{1}&=&(1.32, 1.95), \quad b_{2}=(2.70, 2.05).
\end{array}
\right.
\label{eq:abgen}
\end{equation}
When these solitons are sent to collide head-on, with or without a
$\theta_{1}$ term, the scattering, as usual, takes place at $\pi/2$ radians.
We shall depict this event within the stable format of the modified model.

\begin{figure}
\epsfverbosetrue
\centerline{\epsfbox{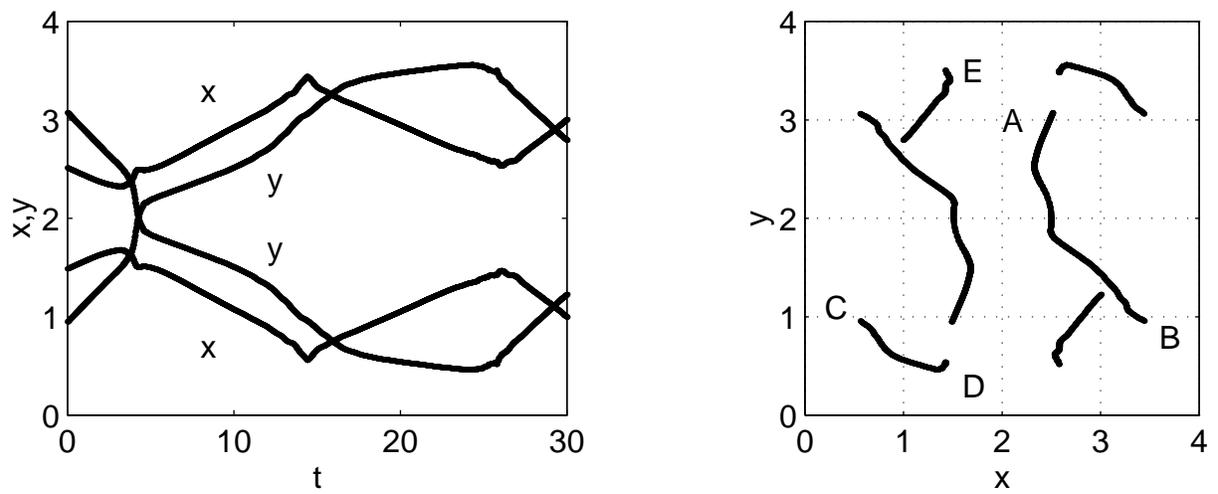}}  
\caption{Trajectories of the position of $Emax$ corresponding
to the head-on scattering for skyrmions arbitrarily situated 
in the basic cell. The labels A-E indicate the itinerary of
one of the lumps.}
\label{fig:gentraj} 
\end{figure}

Figure (\ref{fig:gentraj}) refers to two
skyrmions directed towards each other with $v=(0.18047, 0.0862)$, the speed
being 0.2.  The
coordinates $(x,y)$ correspond to the position of the amplitude $Emax(t)$. 
The labels A-E are a guide as
to the path followed by one of the lumps, the route of the other being given
by the corresponding symmetrical points.
So a skyrmion-lump starts at $A$ and after
$90^{\circ}$ scattering around the centre it continues its itinerary to the
position $B$, where it disappears to re-emerge at $C$. Thence the extended
structure heads south-east and, having reached point $D$ at $t_D \approx$
14.5, it suddenly changes its path to move south-west (point $E$),
unequivocally signalling that a second $90^{\circ}$ scattering has taken
place.  Regarding the other colliding entity, the one starting at \(
\approx (1.5,0.9) \), we can see that the said
second collision changes the trajectory of this skyrmion from the north-west to
the north-east direction.  Our numerical simulation terminates at \( t_F
\approx 30 \), where $F$ denotes the end of the leg started at $E$.

Note that it is also possible to imagine our solitons as evolving on the
surface of a doughnut in ${\cal R}_{3}$, obtained by rotating the circle
of radius $r$ and circumference $L$ (the size of the flat manifold
$T_{2}$) about a coplanar line ($Z$ axis, say) that does not intersect it. The
coordinates $(x \equiv \Upsilon,y \equiv \vartheta)$ serve as the angle of
rotation of the plane of the circle and the angle on the circle itself,
respectively. The parametric equations of such a torus are the standard 
%
\begin{equation}
X=[R+r\cos(\vartheta)]\cos(\Upsilon), \, 
Y=[R+r\cos(\vartheta)]\sin(\Upsilon), \, Z=r\sin(\vartheta).  
\label{eq:terna}
\end{equation} 
Both the radius $r$ and the distance $R$ from the centre of the circle to the
axis of revolution ($Z$) can be calculated from $L$. With the help of
(\ref{eq:terna}), the distance $d$ of a lump on the surface of the torus from
the origin (0,0,0) may be computed via \( d=\sqrt{ X^2+Y^2+Z^2} \).

Finally, regarding the situation  where the initial velocity of the lumps
equals zero, we recall that in the Skyrme model on the sphere the solitons
(represented by an approximate field solution) slighty move away from each
other, showing the presence of a repulsive force between them
\cite{minsk}. On the other hand, on the torus we have found that our
skyrmions, also an approximate solution of the equations of motion, undergo no
translation at all as the time elapses. This agrees with our expectations: The
net repulsive force on a given lump is now zero due to the presence of similar
entities in neighbouring lattices.

\section{Concluding remarks \label{sec:conclusions}}

With the help of numerical simulations we have investigated some
stability and scattering properties of the non-linear $O(3)$ and
Skyrme models with periodic boundary conditions in (2+1) dimensions.

The toroidal $O(3)$ theory has the distinctive feature of possessing analytic
soliton solutions only of degree two and higher. This is because the defining
fields are elliptic functions. We studied a single-soliton case through a
periodic {\em ansatz}, which has turned out to be unstable: After some time
the lump of energy grows too spiky and the numerical procedure breaks down.
Since there are no analytical solutions with topological charge equal to
unity, we may regard the above instability as intrinsinc to the model rather
than an artifact of our numerical method. However, our {\em ansatz} has become
stable upon the addition of an extra term (a Skyrme-like term) to the $O(3)$
Lagrangian and, remarkably, under such circumstances our proposed field serves
as a good, if approximate, soliton solution of degree one on the torus. Note
that unlike the more familiar model on compactified plane, where a
second extra term is required to stabilize the
solitons, our model necessitates only a Skyrme term to achieve so. In this
sense, our model resembles more closely its parent (3+1) dimensional version,
where no second extra term is needed, either.

With regards to collisions we have limited ourselves to those involving two
solitons (analytic solutions of which do exist on $T_{2}$). Scattering at
90$^{\circ}$ was observed in all cases considered, both in the pure $O(3)$ and
Skyrme schemes. Within the framework of the former the lumps shrink unstably
as they evolve, behaviour set off by the perturbation brought about by the
discretization procedure. Such instability is of the same type as that
exhibited in the familiar $O(3)$ model defined on the compactified plane or
topological sphere. As in the one-soliton case, the sole addition of a Skyrme
term stabilizes our two-soliton system. Finally, when the skyrmions start off
from rest they remain motionless under the numerical simulation, in contrast
with the skyrmions on the sphere where they move away from each other due to a
repulsive force amongst them.  

\vspace{10 mm}
\Large{\bf Acknowledgements} \\

\normalsize
 We thank M. Blatter and R. Burkhalter for sending us a subroutine
that computes the Weierstrass $\sigma$ function.  We are also grateful to B.
Piette, J. M. Speight and P. M. Sutcliffe for helpful discussions. RJC is
indebted to {\em Universidad del Zulia} for their financial support.  

\newpage	 

%

\begin{figure}
\epsfverbosetrue
\centerline{\epsfbox{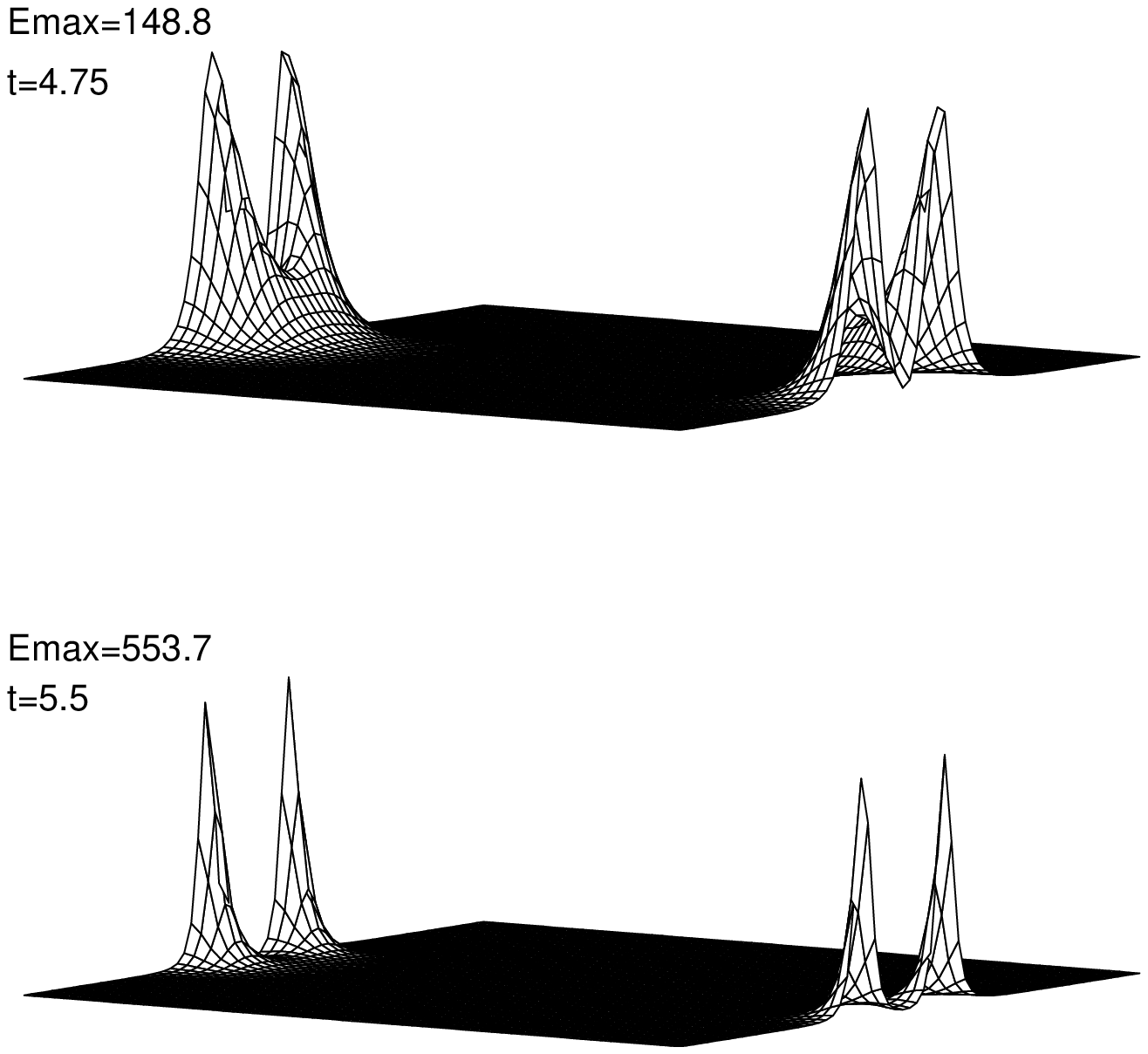}}
\normalsize{Figure \ref{fig:away} (continued): The solitons 
scatter at 90$^{\circ}$. They become very spiky as time 
progresses but, as shown in figure (\ref{fig:conaway}), this is 
corrected by adding a Skyrme term to the Lagrangian.}
\end{figure}

\end{document}